\documentclass[twocolumn,runningheads]{svjour2}
\smartqed  
\usepackage{graphicx}

\journalname{Astrophysics and Space Science}
\begin{document}

\title{Whipple Telescope Observations of LS I +61 303: 2004-2006
}


\author{A. Smith$^{1,12}$ \and R.W. Atkins$^{2}$ \and S. Bradbury$^{1}$ \and O. Celik$^{3}$ \and Y.C.K. Chow$^{3}$ \and P.Cogan$^{4}$ \and C. Dowdall$^{4}$ \and S.J. Fegan$^{3}$ \and P. Fortin$^{5}$ \and D. Gall$^{6}$ \and G.H. Gillanders$^{7}$ \and J. Grube$^{1}$ \and K.J. Gutierrez$^{8}$ \and T.A. Hall$^{9}$ \and D. Hanna$^{10}$ \and J. Holder$^{11}$ \and D. Horan$^{12}$ \and S.B. Hughes$^{8}$ \and T.B. Humensky$^{16}$ \and I. Jung$^{8}$ \and P. Kaaret$^{13}$ \and G. Kenny$^{7}$ \and M. Kertzman$^{18}$ \and D.B. Kieda$^{2}$ \and A. Konopelko$^{6}$ \and H. Krawczynski$^{8}$ \and F. Krennrich$^{13}$ \and M.J. Lang$^{7}$ \and S. Le Bohec$^{2}$ \and G. Maier$^{10}$ \and J. Millis$^{6}$ \and P. Moriarty$^{14}$ \and R.A. Ong$^{3}$ \and  J.S. Perkins$^{12}$ \and K.Ragan$^{10}$ \and G.H. Sembroski$^{6}$ \and J.A. Toner$^{7}$ \and L. Valcarcel$^{10}$ \and V.V. Vassiliev$^{3}$ \and R.G. Wagner$^{15}$ \and S.P. Wakely$^{16}$ \and T.C. Weekes$^{12}$ \and R.J. White$^{1}$ \and D.A. Williams$^{17}$}
\institute{\email{smith@egret.sao.arizona.edu} \and 
 \at (1) School of Physics and Astronomy, University of Leeds (2) Physics Department, University of Utah (3) Department of Physics and Astronomy, UCLA (4) School of Physics, University College Dublin, (5) Department of Physics and Astronomy, Barnard College, Columbia University (6) Department of Physics, Purdue University (7) Physics Department, National University Ireland Galway (8) Department of Physics, Washington University St. Louis (9) University of Arkansas, Dept. of Physics and Astronomy (10) Physics Department, McGill University(11) Department of Physics and Astronomy, University of Delaware (12) Fred Lawrence Whipple Observatory, Harvard-Smithsonian Center for Astrophysics  (13) Department of Physics and Astronomy, Iowa State University  (14) Department of Physical and Life Sciences, Galway-Mayo Institute of Technology (15) Argonne National Laboratory (16) Enrico Fermi Institute, University of Chicago (17) SCIPP and Department of Physics, University of California, Santa Cruz (18) Department of Physics and Astronomy, DePauw University, }

\date{Received: date / Accepted: date}

\maketitle

\begin{abstract}
	In this paper we present the results of the past two years' observations on the galactic microquasar LS I +61 303 with the Whipple 10m gamma-ray telescope. The recent MAGIC detection of the source between 200 GeV and 4 TeV suggests that the source is periodic with very high energy (VHE) gamma-ray emission linked to its orbital cycle. The entire 50-hour data set obtained with Whipple from 2004 to 2006 was analyzed with no reliable detection resulting. The upper limits obtained in the 2005-2006 season covered several of the same epochs as the MAGIC Telescope detections, albeit with lower sensitivity. Upper limits are placed on emission during the orbital phases of 0$\rightarrow$0.1 and 0.8$\rightarrow$1, phases which are not included in the MAGIC data set. 

\end{abstract}

\section{LS I +61 303 as a VHE Emitter}
\label{sec:1}

	LS I +61 303 is a high-mass X-ray binary system located at a distance of roughly 2kpc \cite{Ref6}. The system is comprised of a compact object, most likely a neutron star (although the presence of a black hole has not been ruled out), in a highly eccentric orbit around a rapidly rotating B0-B0.5 main sequence star \cite{Ref7}. The most accurate determination of the orbital parameters of the system comes from optical spectroscopy measurements \cite{Ref6}. In that paper, the authors use the period of 26.496 $\pm$ 0.0028  days \cite{Ref8} along with their optical measurements to derive an orbital phase of $\phi$= 0.22 $\pm$ 0.02 for periastron passage (see figure 1). From the high energy perspective, interest in this system began in 1978 \cite{Ref9},  when the authors proposed the association of LS I +61 303 with the COS B gamma-ray source 2CG 135+01.  Later, it would also be associated with the EGRET unidentified source 3EG J0241+6103 \cite{Ref4}. The identification of LS I +61 303 as a microquasar occurred in 2001 \cite{Ref3} when relativistic, precessing radio jets were discovered extending roughly 200 AU from the center of the source.

\begin{figure}
\begin{center}
  \includegraphics[width=75mm,height=70mm]{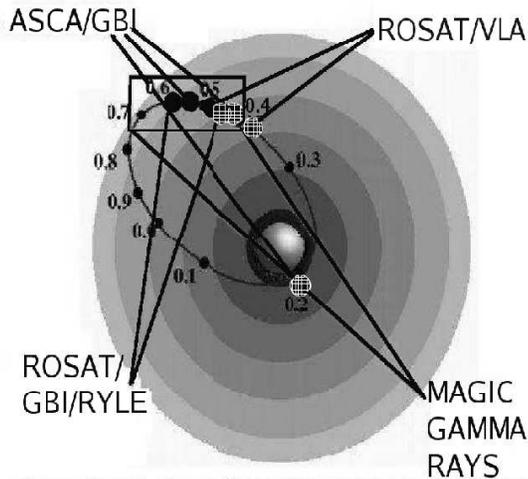}
\end{center}
\caption{The elliptical orbit of the compact object in LS I +61 303 and associated radio, X-Ray, and TeV observations in terms of its orbital phase. The hashed circles represent X-ray flares, whereas the solid circles represent radio flares. Modified from \cite{Ref11}.}
\label{fig:1}       
\end{figure}
	An intriguing feature of this system is the presence of periodic modulation of outbursts in both radio and X-ray frequencies on a time scale of roughly 26 days, coincident with the orbital period of the system \cite{Ref6}. In the case of the radio emission, the amplitude and the phase of these outbursts is also modulated, with a period of roughly 4.6 years \cite{Ref8}. The exact mechanism for this long term modulation is unclear, but it is believed to be linked to periodic changes in the stellar wind structure of the Be star. Multi-wavelength campaigns undertaken by ASCA, ROSAT, VLA, GBI and RYLE  indicate that most hard X-ray and radio emission is localized between $\phi$ = 0.35 and 0.65 (\cite{Ref1}\cite{Ref2}\cite{Ref10}); however, radio outbursts have been detected all the way out to $\phi \approx$ 0.9 \cite{Ref8}. The persistent feature of these observations is that hard X-ray outbursts tend to precede the radio outbursts by a few days. It has been suggested \cite{Ref11} that the most likely physical mechanism for this correlation is the increased disk accretion rate (resulting in a spectral shift in X-rays to a hard state) which then fuels radio-loud jet ejection events. It is asserted that if these events take place in a region of the orbit where a large fraction of stellar UV photons is present, gamma-ray emission can take place via inverse Compton scattering of energetic jet electrons off this photon population. Recently, the MAGIC collaboration has detected VHE gamma-ray emission between $\phi$= 0.4 and 0.7, which is consistent with the previous multi-wavelength results \cite{Ref5}. 

If we assume that the VHE emission in this system is definitely linked to accretion increases in the disk (fueling jets that scatter UV photons), we should look for regions of the orbit where jet production can take place and stellar UV photon exposure is readily available. There are two regions in the orbit in which this is believed to be possible \cite{Ref11}, the first at periastron passage where the UV photon and stellar wind densities are the highest. The second region occurs at a point in the orbit where the decrease in orbital velocity with respect to the stellar wind compensates for the decreased stellar wind density ($\frac{dM}{dt} \propto \frac{\rho}{v^{3}}$), resulting in a second accretion rate peak ($\approx \phi$= 0.5 $\rightarrow$ 0.8) \cite{Ref14}. If, during this second accretion peak, the jet is exposed to a significant population of stellar UV photons, then it is expected that VHE gamma-ray emission should take place. 

It is during and around these expected accretion peaks that we would expect to see radio, X-ray, and gamma-ray emission from LS I +61 303. This assertion is given credence by \cite{Ref4}, in which the authors detail EGRET observations showing significant gamma-ray flux above 100 MeV both at periastron and around $\phi$ = 0.5. However, this conflicts with the fact that the MAGIC data shows no VHE emission around periastron. Morever, no radio outbursts have ever been detected around periastron. This may indicate that there some secondary effect inhibits or obscures emission near periastron passage, but the EGRET detection contradicts this. If it is still to be maintained that the broadband emission from this system can be explained as resulting from accretion powered jets, these observational discrepancies must be addressed. 

There has also been an alternate model for VHE emission from LS I +61 303 which postulates that the system is not a microquasar at all, but rather a binary pulsar system \cite{Ref15}. In this model, the gamma-ray emission results from stellar photons upscattered by electrons originating from the rotationally powered pulsar wind. Although binary pulsar systems (such as PSR 1259-63) have been detected in TeV gamma-rays, it remains unclear whether the current observational data on LS I +61 303 can support such a hypothesis. After decades of observations on LS I +61 303, the exact nature of emission from this system remains unknown.

\section{The Whipple Gamma-Ray Telescope}
\label{sec:2}

	The Whipple Observatory 10m Gamma-Ray Telescope located on Mt. Hopkins in southern Arizona (2300 m altitude) was built in 1968 and was the first imaging Cherenkov telescope in gamma-ray astronomy. Sensitive in the energy range from 350 GeV to roughly 10 TeV, the Whipple telescope consists of 248 individual hexagonal mirror facets with a total area of approximately 75 m$^{2}$. Pulses of Cherenkov light from the air showers initiated by high-energy gamma rays entering the Earth's atmosphere will form pools of light at ground level with a typical radius of 120 m. If just part of this pool of light can be sampled, the air shower can be detected; thus the total effective area of the Whipple telescope is roughly 10$^{4}$ $\rightarrow$ 10$^{5}$ m$^{2}$ (depending on the energy of the gamma ray which initiated the shower).

\begin{figure}
\begin{center}
  \includegraphics[width=75mm,height=70mm]{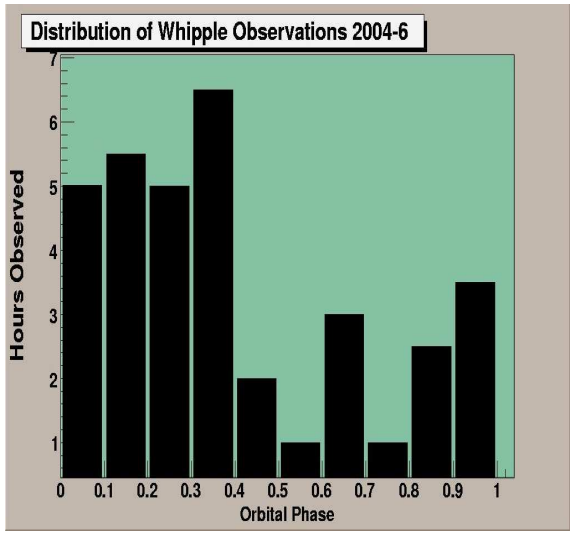}
\end{center}
\caption{Distribution of the hours observed during each orbital phase from 2004-2006.}
\label{fig:2}       
\end{figure}

The telescope images the Cherenkov light produced by air showers on a camera consisting of an array of 379 PMTs (pixels). The pulse amplitude from each pixel is measured using charge-amplitude converters with an integration time of 30 ns. A topological pattern trigger looks to see if pixels that are grouped together fired within a set time window. At this stage, the system is firing triggers on light coming from both gamma-ray induced showers and background noise. This background noise consists mainly of air showers initiated by charged cosmic rays, but night-sky background noise and local muons also contribute to this noise (these two are more significant at lower energies). These background signals represent a significant hurdle to any reliable detection of a cosmic gamma-ray source, as their isotropic presence is typically several orders of magnitude stronger than any gamma-ray source. This problem is overcome by utilizing the intrinsic differences between the development of air showers initiated by gamma rays and those initiated by hadrons. The key distinction is that images of gamma-ray showers have smaller angular spread and are more tightly bunched in time. The separation of background showers occurs at the software level with code developed to distinguish pixel firing patterns that are inconsistent with the geometry expected for a gamma-ray induced shower. This selection process eliminates $\>$99$\%$ of the background noise, while retaining $\>$50$\%$ of the desired gamma-ray triggers. After all levels of data cleaning, the Whipple telescope is capable of detecting point sources with $\approx$5$\%$ of the Crab Nebula flux at the 5$\sigma$ level in 50 hours. 

\begin{figure}
$\begin{left}
  .\includegraphics [width=75mm,height=70mm] {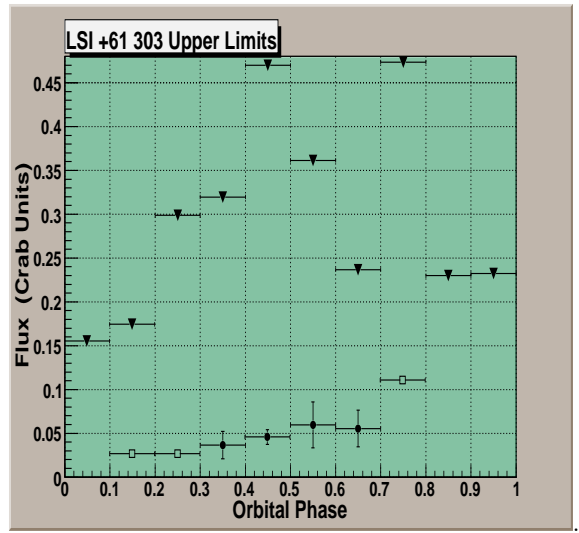}
.\right .\end{left}$
\caption{ Shown here are the 99.9$\%$ upper limits in terms of Crab flux calculated by orbital phase (triangles). Shown for comparison are the MAGIC detections (points with errors) and 90$\%$ upper limits (boxes) reported in \cite{Ref5}.}
\label{fig:3}       
\end{figure}

\section{Observational Results}
\label{sec:3}

	From September 2004 to February 2006, a total of 50 hours of observations were taken on LS I +61 303. Of these 50 hours, roughly 40 hours had accompanying control observations which were used to determine the background for the observations. A set of cuts were then applied to the data, eliminating runs taken under poor weather or runs with erratic raw trigger rates. After this process, 35 hours remained of the original data set, and it was this reduced set that was used for the analysis presented here.  Cuts based on standard parameterization techniques [12] were applied for gamma-hadron separation.

 	All orbital phases were covered with these observations, though not with equal exposure time because of the interruption in observations caused by moonlight. It is unfortunate that the phases with relatively small data coverage were some of the phases during which MAGIC saw the largest excess of gamma-ray activity (see figure 2). From 2004-2006, the Whipple data showed no evidence for a detection. Of the six epochs which provided the MAGIC detection, Whipple had four concurrent observation periods with no similar strong detections.

 	Since there were no significant detections, the data were used to determine 99.9$\%$ flux upper limits by the method described in \cite{Ref16}. Upper limits were derived for all orbital phases and are shown in figure 3 in terms of flux from the Crab Nebula during the same observational period. While the limits derived for the orbital phases of 0.1$\rightarrow$0.8 are less stringent than the ones derived in [5] the MAGIC result did not include any coverage in the phase bins of 0$\rightarrow$0.1, and 0.8$\rightarrow$1. The Whipple upper limits for these phase bins therefore serve as a constraint on the system emission in the phases in which it is most likely quiescent.  Finally, a total flux upper limit of 0.072 Crab units was derived for flux above 350 GeV. 

\begin{figure}
$\begin{left}
  .\includegraphics[width=75mm,height=70mm]{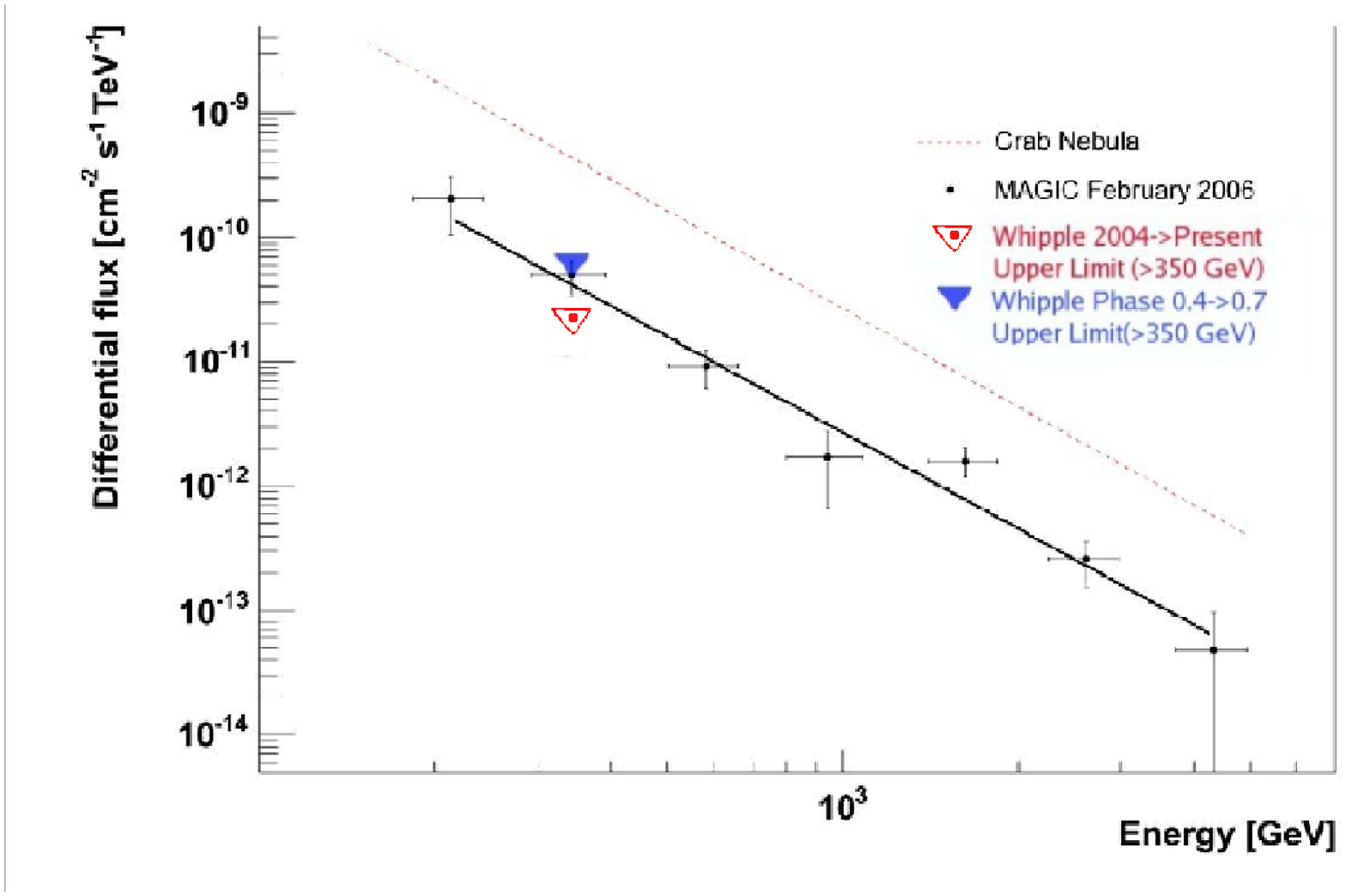}
.\right .\end{left}$
\caption{The comparison between the MAGIC extracted spectrum (line) and the upper limit for emission derived from Whipple observations. The integral flux upper limit points obtained were translated (via the MAGIC Crab nebula spectrum) to the upper limit points shown in the graph. (figure modified from \cite{Ref5}).}
\label{fig:4}       
\end{figure}

\begin{table}[t]
\caption{2004$\rightarrow$2006 LS I +61 303 observations binned by orbital phase.}
\centering
\label{tab:1}       
\begin{tabular}{llll}
\hline\noalign{\smallskip}
Orbital Phase & On Source & Off Source & Minutes\\[2pt]
 & Counts & Counts & On Source \\[2pt]
\tableheadseprule\noalign{\smallskip}
0.0$\rightarrow$0.1 & 1865 & 1965 & 275 \\ 
0.1$\rightarrow$0.2 & 2117 & 2187 & 302\\ 
0.2$\rightarrow$0.3 & 1969 & 1924 & 275\\ 
0.3$\rightarrow$0.4 & 2608 & 2503 & 357\\ 
0.4$\rightarrow$0.5 & \enspace 739 & 707 & 110\\ 
0.5$\rightarrow$0.6 & \enspace 264 & 284 &\enspace 55 \\ 
0.6$\rightarrow$0.7 & 1097 & 1142 & 165 \\ 
0.7$\rightarrow$0.8 & \enspace 374 & 387 & \enspace 55 \\ 
0.8$\rightarrow$0.9 & \enspace 751 & 791 & 137 \\ 
0.9$\rightarrow$1 & 1174 & 1202 & 192\\
\noalign{\smallskip}\hline
\end{tabular}
\end{table}

When plotted against the MAGIC spectrum (dotted triangle in figure 4), it would seem that this result is in conflict with that result; however, the MAGIC spectrum is derived from observations taken between 0.4$\rightarrow$0.7 and is not representative of the flux during all orbital phases. A more useful quantity to compare is the flux upper limit for the Whipple data set for the phase periods of $\phi$ = 0.4$\rightarrow$0.7 only: a weaker limit of 0.184 Crab units for flux above 350 GeV (solid triangle in figure 4). It appears that for the limited exposure time ($\approx$ 5 hours) during the active phases of the system, the Whipple telescope was not sensitive enough to detect LS I +61 303.

\section{Future Observations}
\label{sec:4}

	The Whipple telescope will continue to monitor LS I +61 303 in the 2006-7 observing season as part of a planned multiwavelength campaign. The VERITAS collaboration will also collect data on this target with the VERITAS array located at the base of Mt. Hopkins, Arizona. This much more sensitive instrument consists of 4 12m diameter telescopes and is sensitive at the level of detecting a 5$\%$ Crab signal at the 5$\sigma$ level in approximately one hour. 

	Moreover, fall 2006 will be a unique time in terms of observing LS I +61 303. There will be six orbital cycles between September 2006 and February 2007 when
 	LS I +61 303 will be at an acceptable elevation for observation. Since both VERITAS and Whipple
are limited by not being able to observe during moonlight, this constrains our ability to give 
equal coverage time to all phases of orbit. However, the phases of the moon 
which will allow observations coincide well with the orbital phases of $\phi =$ 0.3 $\rightarrow$ 0.75, exactly the regions in which MAGIC detected VHE emission from this object (see figure 5). The 2006-2007 season will also allow for observations of the orbital phases in which emission is not entirely expected ($\phi$ = 0.8 $\rightarrow$ 1), thus placing heavier constraints on emission models. This fall will be an ideal time for both Whipple and VERITAS to extend the VHE observations on this exciting object.

\begin{figure}
$\begin{left}
  .\includegraphics[width=75mm,height=70mm]{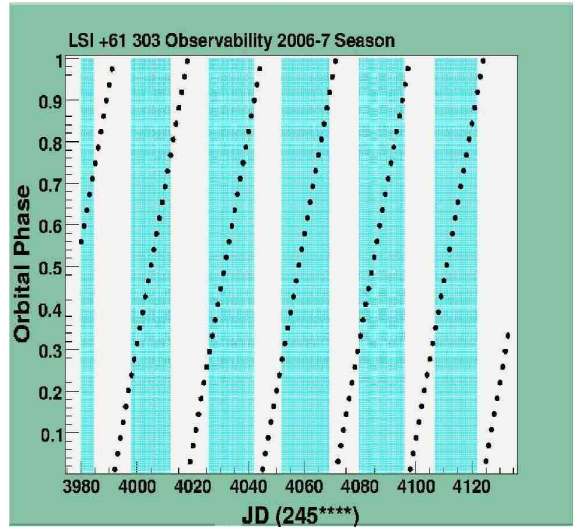}
\right .\end{left}$
\caption{Orbital phases of LS I +61 303 (dots) as a function of  Julian date. The shaded areas are the times that moon phase will allow more than 3 hours of observations per night. A serendipitous agreement is evident between the upcoming schedule of ``dark'' time and the times when LS I +61 303 that are expected to be the most active.}
\label{fig:5}       
\end{figure}

\end{document}